\documentstyle[12pt]{article}

  \newcommand*\ti[5]{{\em #5}, {#1} {\bf #2}, #3 (#4)}

\newcommand*\jhep{JHEP}
\newcommand*\np{Nucl. Phys.}
\newcommand*\pl{Phys. Lett.}

\if@twoside  m
\else
\fi
 1
 1
\font\sr = msbm10 scaled \magstep 1

\def\BC{\mbox{\sr C}}
\def\BR{\mbox{\sr R}}

\def\beq{\begin{equation}}
\def\eeq{\end{equation}}
\def\beqa{\begin{eqnarray}}
\def\eeqa{\end{eqnarray}}

\begin{document}
\begin{titlepage}
\rightline{MPI}
\rightline{LMU-TPW 2000-09}
\vspace{4em}
\begin{center}

 {\LARGE{\bf Noncommutative gauge theory for Poisson manifolds}}

\vskip 1.5cm

{{\bf Branislav Jur\v co,${}^{*}$ Peter Schupp${}^{**}$ and Julius Wess${}^{**}$ }}

 \vskip 0.5 cm

${}^*$Max-Planck-Institut f\"ur Mathematik\\Vivatgasse 7\\
D-53111 Bonn, Germany\\[1ex]
${}^{**}$Sektion Physik\\
Universit\"at M\"unchen\\
Theresienstr.\ 37\\
D-80333 M\"unchen, Germany

 \end{center}

 \vspace{1 cm}

 \begin{abstract} 
A noncommutative gauge
theory is associated to every Abelian gauge theory on a Poisson manifold.
The semi-classical and full quantum version of the map from the ordinary gauge
theory to the noncommutative gauge theory (Seiberg-Witten map) is given
explicitly to all orders for any Poisson manifold in the Abelian case. 
In the quantum case the 
construction is based on Kontsevich's formality theorem.
\end{abstract}
\vfill
\hrule
\noindent
\small{\it e-mail: }\\
{\small\quad jurco@mpim-bonn.mpg.de}\\
{\small\quad schupp@theorie.physik.uni-muenchen.de}\\{\small \quad wess@theorie.physik.uni-muenchen.de}
\end{titlepage}\vskip.2cm

\newpage

\setcounter{page}{1}
\section{Introduction}
Noncommutative geometry naturally enters the description of open strings in a
background $B$-field \cite{CLNY,CH,S}. The D-brane world volume is then
a noncommutative space whose fluctuations are governed by a noncommutative
version of Yang-Mills theory \cite{CDS,DH,MZ,W,SW}. 
In the case of a constant $B$-field it has been
argued that there is an equivalent description in terms of ordinary gauge
theory. From the physics perspective the two pictures are related by a choice of
regularization~\cite{SW,AD}. There must therefore exist a field redefinition -- a
Seiberg-Witten map~\cite{SW}. The $B$-field, if non-degenerate and closed,
defines a symplectic structure on the D-brane world volume; its inverse is a
Poisson structure whose quantization gives rise to the noncommutativity.

An interesting question arises: Given a gauge theory on a general Poisson
manifold -- is there always a corresponding noncommutative gauge theory on the
noncommutative space -- the quantization of the original Poisson manifold?
Previously we found this to be true for symplectic manifolds~\cite{JS}.%
\footnote{For a detailed description of the universal gauge theory of the Weyl-Bundle
see, e.g.,~\cite{AK}.}
Here we give the construction for an arbitrary Poisson manifold. (The present discussion
is much more explicit and complete.) On the way an appropriate generalization of
Moser's lemma from symplectic geometry to the Poisson case and its quantization
are given. This, we believe, is mathematically interesting in its own right.
As in our previous paper~\cite{JS}, we choose to work within the framework of
deformation quantization~\cite{BFFLS,Kontsevich,Sternheimer}. 
This allows us to postpone questions related to
representation theory, so that we can focus on the algebra. 
We expect that our results can be used to find derivative corrections to the 
Born-Infeld action, classifying invariant actions in the spirit of~\cite{C2}.


\section{Noncommutative Yang-Mills theory}

Here we recall how the gauge theory on a more-less arbitrary noncommutative 
space was introduced in \cite{Wess}. 
The formulation starts with an associative, not necessarily commutative, 
algebra ${\cal A}_x$ over $\BC$ freely generated by finitely many generators 
$\hat{x}^i$ modulo some relations $\cal R$.  ${\cal A}_x$ plays role of the 
noncommutative space-time.
The matter fields $\psi$ of the theory
are taken to be elements of a left module of  ${\cal A}_x$ and the infinitesimal
gauge transformation induced by $\hat{\lambda} \in {\cal A}_x$ is 
given by the left multiplication (action)
\beq
\psi \stackrel{(\hat\lambda)}{\mapsto} \psi + i\hat{\lambda} \psi.\label{ncgt1}
\eeq
The gauge transformation does not act on the ``coordinates" $\hat{x}^i$.
\beq	 
\hat{x}^i \stackrel{(\hat {\lambda})}{\mapsto} \hat{x}^i. \label{ncgt2} 
\eeq
The left multiplication of a field by the coordinates  $\hat{x}^i$ is not covariant under the gauge transformation
\beq
\hat{x}^i\psi \stackrel{(\hat {\lambda})}{\mapsto} \hat{x}^i\psi+i\hat{x}^i \hat{\lambda} \psi,
\eeq
since in general $ \hat{x}^i\hat{\lambda} \psi$ is not equal to $\hat{\lambda} \hat{x}^i \psi$.
The gauge fields $\hat{A}^i$, elements of ${\cal A}_x$, are introduced to cure this. Namely, covariant coordinates 
\beq\hat{X}^i = \hat{x}^i +  \hat{A}^i 
\eeq
are introduced. 

The gauge transformation is supposed to act on the gauge
fields  $\hat{A}^i$ in a way that will assure the covariance of $\hat{X}^i \psi$
under the gauge transformation (\ref{ncgt1}),(\ref{ncgt2}).
This is achieved by the prescription
\beq
\hat{A}^i \stackrel{(\hat {\lambda})}{\mapsto} \hat{A}^i + i[\hat{\lambda},  \hat{x}^i] + i[\hat{\lambda},  \hat{A}^i].
\eeq
In examples considered in \cite{Wess} (universal enveloping algebra of a finite
dimensional Lie algebra and of the Heisenberg algebra as a special case,
quantum plane)  also the corresponding field strength
$\hat{F}^{ij}$ was introduced. If 
\beq
[\hat{x}^i,\hat{x}^j] = J^{ij}(\hat x),
\eeq
then  
\beq
\hat{F}^{ij}=[\hat{X}^i,\hat{X}^j] -   J^{ij}(\hat X).
\eeq
Of course this is not unique, the choice of ordering in the above formula may effect the definition
of $\hat F$, but this is not important for covariance.
It follows
\beq
\hat{F}^{ij} \stackrel{(\hat {\lambda})}{\mapsto}\hat{F}^{ij} + i[\hat{\lambda}, \hat{F}^{ij}],
\eeq
as expected.

This construction covers also the noncommutative generalization of non-Abelian
($GL(N)$ or $U(N)$ if ${\cal A}_x$ possesses a $\ast$--structure) gauge theories, taking
the tensor product algebra ${\cal A}_x\otimes U(gl(N))$ instead of ${\cal A}_x$.

The following question inspired by \cite{SW} appears naturally.

Let us assume that our associative algebra ${\cal A}_x$  (and this was indeed
the case of the examples considered in \cite{Wess}) can be understood
as a deformation
quantization of a commutative algebra of functions on some Poisson manifold $M$.
Let $\star$ be the corresponding star product.
Let us also assume that we have a (non-Abelian) gauge field $A$ on $M$.

Does there exist a map $SW$ :
\beq
A \stackrel{(SW)}{\mapsto} \hat A, \hskip 1cm 
\lambda \stackrel{(SW)}{\mapsto} \hat{\lambda}(\lambda, A) 
\eeq
such that the (non-Abelian) commutative gauge transformation on $A$
\beq
A \stackrel{(\lambda)}{\mapsto} A + d\lambda + i[\lambda, A] \label{clg}
\eeq
is sent by $SW$
into the noncommutative gauge transformation on $\hat A$
\beq
\hat{A}^i \stackrel{(\hat{\lambda)}}{\mapsto} \hat{A}^i + i[\hat{\lambda},  \hat{x}^i]_{\star}
 + i[\hat\lambda,\hat{A}^i]_{\star} \label{qg}\hskip 2mm ?
\eeq

The commutator in (\ref{clg}) is the matrix one and the commutator in (\ref{qg}) is the star
commutator on functions and the matrix one on matrices.

In this paper we give a general and explicit construction of the map $SW$ in the Abelian case.
Our main tool is Kontsevich's formality theorem. 

The non-Abelian case is more involved and will be treated by similar methods in the sequel to this paper.

\section{Classical}
Here we formulate the classical analogue of the Seiberg-Witten map between the
commutative and noncommutative description of Yang-Mills theory for any Poisson
manifold.

Let $M$ be a manifold and $F$ a two-form on $M$.
Let us temporarily assume that $F$ is exact. Later on we will relax this condition,
$F$ closed will appear to be good enough for our purposes.
In local coordinates we write $A=A_i d x^i$ and $F=F_{ij} d x^i\wedge d x^j$, with
$F_{ij}=\partial_i A_j -\partial_j A_i$.
Let us further assume that we have on $M$ a one-parameter family
of bivector fields $\theta(t) =\frac{1}{2}\theta^{ij}(t)\partial_i\wedge\partial_j$, $t\in[0,1]$,
with the explicit $t$-dependence of the matrix $\theta^{ij}(t)$ given
by
\beq
\partial_t\theta(t)=-\theta(t) F \theta(t), \label{time}
\eeq
with the initial condition
\beq
\theta(0)=\theta,
\eeq
where $\theta$ is some fixed but otherwise arbitrary Poisson tensor on $M$.
The product on the right in (\ref{time}) is the matrix one. As above
we will often use the same notation for polyvector fields or forms and the corresponding 
tensors.
The formal solution to (\ref{time}) can be given by the following power series
in $t$ (or in $\theta$)
\beq
\theta(t)=\sum_{n\geq 0} (-t)^n \theta(F\theta)^n \label{sol}.
\eeq
The convergence is not an issue here, because we will work all the time
with formal power series in $\theta$. (E.g. if the matrix $\theta$ is invertible then 
in the physical situation of a $D$-brane in the background of the $B$-field
$B=\theta_{ij}^{-1}d x^i \wedge d x^j$ it simply means that the background is strong.)

It follows from (\ref{time}), or directly from (\ref{sol}), that $\theta(t)$ continues to be a Poisson tensor
for all $t\in [0,1]$. For this only the closedness of $F$ is important.
The Poisson bivector field $\theta(t)$ defines an bundle map $T^*M 
\rightarrow TM$ given by $i_{\theta(t) (\omega)} \eta = \theta(t)(\omega, \eta)$ for any 
one-forms $\omega$ and $\eta$. 
Using the Jacobi identity $[\theta(t), \theta(t)]=0$, with $[ , ]$ being the Schouten-Nijenhuis 
bracket of polyvector fields, we can easy verify the $t$-derivative of $\theta(t)$
is given by a Lie-derivative
that 
\beq
\partial_t \theta(t) + [\chi(t), \theta(t)]=0 ,\label{Lie}
\eeq
where now $\theta(t)$ is understood as a bivector field
 and 
\beq \chi(t)= \theta(t) (A)
\eeq
 is a vector field that in local coordinates looks like
$$
\chi(t)=\theta^{ij}(t)A_i\partial_j.
$$
Let us recall that the Schouten-Nijenhuis bracket of two polyvector fields is defined
by
$$
[\xi_1\wedge...\wedge\xi_k,\eta_1\wedge...\wedge\eta_l]=
\sum_{i=1}^{k}\sum_{j=1}^{l}(-1)^{i+j}
[\xi_i,\eta_j]\wedge\xi_1\wedge...\wedge\hat{\xi}_i\wedge...\wedge
\hat{\eta}_j\wedge...\wedge\eta_l
$$
$$
[\xi_1\wedge...\wedge\xi_k,f]=\sum_{i=1}^{k}(-1)^{i-1}\xi_i(f)\xi_1\wedge...\wedge\hat{\xi}_i
\wedge...\wedge\xi_k,
$$
if all $\xi$'s and $\eta$'s are vector fields and $f$ is a function.

If $f$ and $g$ are two smooth functions on $M$ with no explicit dependence on $t$ and 
$\{ , \}_t$ denotes the Poisson bracket corresponding to $\theta(t)$ then (\ref{Lie}) is 
rewritten as
\beq
\partial_t \{f,g\}_t + \chi(t)\{f,g\}_t -\{\chi(t)f,g\}_t -\{f,\chi(t)g\}_t =0. \label{der}
\eeq
Both (\ref{Lie}) and (\ref{der}) imply that all the Poisson structures $\theta(t)$ are related 
by the flow $\rho^*_{tt'}$ of $\chi(t)$:
$\rho^*_{tt'}\theta(t')=\theta(t)$.
Setting
$\rho^*=\rho^*_{01}$ we have in particular
\beq
\rho^*\theta' = \theta,
\eeq
i.e.,
\beq
\rho^* \{f,g\}' = \{\rho^* f, \rho^* g\},
\eeq
where $\theta'$ is short for $\theta(1)$.
The vector field $\chi(t)$ may not be complete, however
$\rho^*$ again has to be understood as a formal 
diffeomorphism given by formal power series in 
$\theta$. In this sense we always have
a (formal) coordinate change on $M$
which relates the two Poisson structures $\theta$ and $\theta'$.
Explicitly
\beq
\rho^* = \left.e^{\partial_t + \chi(t)}e^{-\partial_t}\right |_{t=0}.
\label{exp}  
\eeq

Consider now a gauge transformation 
\beq
A\mapsto A +d\lambda.\label{gauge}
\eeq
The effect upon $\chi(t)$ will be
\beq
\chi(t)\mapsto \chi(t)+\chi_{\lambda}(t),\label{gaugeonfields}
\eeq
where $\chi_{\lambda}(t)$ is the Hamiltonian vector field
\beq
\chi_{\lambda}(t)=\theta(t)(d\lambda)=[\theta(t),\lambda]\eeq
and $[\chi_{\lambda}(t), \theta(t)]= 0$. In local coordinates 
$\chi_{\lambda}=\theta^{ij}(t)
(\partial_i\lambda)\partial_j$.
Correspondingly we use the notation $\rho^*_{\lambda,tt'}$ for the new flow.
It follows almost immediately that
$\rho^*_{\lambda}(\rho^*)^{-1}=\left.e^{\partial_t + \chi_{\lambda}(t)}e^{-\partial_t - 
\chi(t)}\right|_{t=0}$
is generated by a Hamiltonian vector field $\theta (d\tilde{\lambda})$
for some $\tilde{\lambda}$. This follows from the Baker-Campbell-Hausdorff identity and the 
fact that
$$
[\partial_t+\theta(t)(A), \theta(t) (df)] = 
\theta (t)(dg)  
$$
with $g= \theta(t)(d\lambda, A)$. Having this in mind it is easy to see all the terms coming from 
B--C--H formula contain only commutators of this type or commutators of two Hamiltonian vector 
fields which are again Hamiltonian ones.
Even more is true: $\rho^*_{{\lambda}, tt'}(\rho^*_{t't})^{-1}$ for all $t$ and $t'$ is 
generated by some Hamiltonian vector field for $\theta(t)$. 

So the transformation induced by $\lambda$ takes the form
\beq
f \stackrel{(\lambda)}{\mapsto} f + \{\tilde\lambda,f\}.\label{tilde}
\eeq 

It is clear from the above discussion that working only with formal power series in $\theta$
we can abandon the exactness condition for $F$ and assume $F$ only closed with the following
consequences. The gauge field $A$ and the vector field $\chi(t)$ are given only locally.
If $A$'s given in two different local patches are related on their intersection by the gauge transformation
(\ref{gauge}), then the corresponding local vector fields are related by 
(\ref{gaugeonfields}) and the local diffeomorphisms $\rho$ are related by 
the canonical transformation (\ref{tilde}) generated by $\tilde{\lambda}$.

In the case of invertible $\theta(t)$ and of a compact manifold we have the well know
lemma of Moser \cite{Moser}.

Let us return to the $SW$ map in classical setting.
For this we have to choose some local coordinates $x^i$ on $M$.
Let us write the result of acting by the diffeomorphism $\rho^*$ 
on the coordinate function $x^i$ in the form \cite{Ishibashi,Okuyama,Cornalba,JS,Wess}.
\beq 
\rho^*(x^i)=x^i + A_{\rho}^i.\label{xplusa}
\eeq
$A_{\rho}$ depends as a formal power series in $\theta$ on $A$.
Explicitly we have
\beq
A_{\rho}^i = (\left.e^{\partial_t + \theta^{ij}(t)A_i\partial_j}-1)\right|_{t=0} x^i 
\eeq
Let us act by the infinitesimal gauge transformation 
(\ref{gauge}) on $A$. This induces the infinitesimal Poisson map 
(\ref{tilde}) on $\rho^*(x^i)$, which in turn induces a map on $A_{\rho}$
given by
\beq
A_{\rho}^i \stackrel{(\lambda)}{\mapsto} A_{\rho}^i + \{\tilde{\lambda},x^i\} +\{\tilde{\lambda}, A^i_{\rho}\}.
\eeq
So the map $A\mapsto A_{\rho}$ can be viewed as the semi-classical version of the 
SW map which we are looking for.

\section{Formality}
The existence of a star product on an arbitrary Poisson manifold follows
from the more general formality theorem~\cite{Kontsevich}: 
There exists an $L_{\infty}$-morphism from the differential graded algebra 
of polyvector fields  into the 
differential graded algebra of polydifferential operators on $M$.
There is a canonical way to extract a star product $\star$ 
from such an $L_{\infty}$-morphism for 
any formal Poisson bivector field. 
We will refer to this star product as the Kontsevich star product.
Any star product on $M$ is equivalent to some  Kontsevich star product.

The differential graded algebra $T_{poly}(M)$ is 
the graded
algebra of polyvector fields on $M$
$$ T^n_{poly}(M)=\Gamma(M,\wedge^{n+1}TM),\,\,n\geq -1,$$ 
equipped with the standard Schouten-Nijenhuis bracket and differential $d\equiv 0$.
An $m$-differential operator in $D_{poly}(M)$ acts on a tensor product of $m$ functions
and has degree $m-1$.  The composition $\circ$ on $D_{poly}(M)$ is given by
\begin{eqnarray*}
(\Phi_1\circ \Phi_2)(f_0 \otimes...\otimes f_{k_1+k_2})=&\\
\sum_{i=0}^{k_1}
(-1)^{k_2 i}\Phi_1 (f_0 \otimes...\otimes f_{i-1}\otimes &(\Phi_2(f_{i} 
\otimes...\otimes f_{i+k_2}))\otimes f_{i+k_2+1}\otimes...\otimes f_{k_1+k_2})
\end{eqnarray*}
for $\Phi_i \in D^{k_i}_{poly}(M)$
and the Gerstenhaber bracket $[\Phi_1,\Phi_2]$ is then given by
\beq
[\Phi_1,\Phi_2]=\Phi_1\circ \Phi_2 - (-1)^{k_1k_2}\Phi_2\circ \Phi_1.
\eeq
The differential on $D_{poly}(M)$ is given in terms of the Gerstenhaber bracket as
\beq d\Phi=-[\mu,\Phi]
\eeq
where $\mu$ is the multiplication of functions: $\mu(f_1\otimes f_2)=f_1f_2$.

An $L_\infty$-morphism $U:T_{poly}(M)\rightarrow D_{poly}(M)$ is then a collection of 
skew--symmetric multilinear maps $U_n$ from tensor products of $n\geq 1$ polyvector fields 
to polydifferential operators of degree $m \geq 0$: 
$\otimes^n T_{poly}(M)\rightarrow D^{m+1}_{poly}(M)$,
satisfying the following condition (formality equation)~\cite{Kontsevich,Arnal}:
\begin{eqnarray} 
\lefteqn{Q_1'U_n(\alpha_1,...,\alpha_n) 
+\frac{1}{2}
\sum_{I\sqcup J=(1,...,n)\atop{I,J \neq \emptyset}}
\epsilon_{\alpha}(I,J)Q_2'(U_{|I|}(\alpha_I),
U_{|J|}(\alpha_J))}&& \label{formality}\\
&&= \frac{1}{2}\sum_{i\neq j}\epsilon_{\alpha}(i,j,1,...,\widehat{i,j},...,n)
(U_{n-1}Q_2(\alpha_i,\alpha_j)\nonumber,
\alpha_1,...,\hat{\alpha}_i,...,\hat{\alpha}_j,...,\alpha_n))
\end{eqnarray}
Here
$Q_1'(\Phi)=[\Phi,\mu]$, $Q_2'(\Phi_1, \Phi_2)=(-1)^{|1|(|2|-1)}[\Phi_1, \Phi_2]$, with $|i|$ 
denoting the degrees of homogeneous polydifferential operators $\Phi_i$ and 
$Q_2(\alpha_1,\alpha_2)=-(-1)^{k_1(k_2+1)}[\alpha_2,\alpha_1]$, where $k_i$ are degrees of 
homogeneous polyvector fields $\alpha_i$. Further $|I|$ denotes the numbers of elements of $I$
and $\epsilon_{\alpha}(I,J)$ is $+1$ or $-1$ depending on the number of
transpositions of odd elements in the permutation of $\{1,...,n\}$ associated with the partition $(I,J)$.
Although it doesn't explicitly enter the formality condition (\ref{formality}) a zero component
$U_0$ can be added to $U$. By definition it is nonzero only if acting on two 
functions $f\otimes g$ with the result $U_0(f,g)=fg$. 

In the case of $M=\BR^d$ Kontsevich gives also a beautiful explicit expression for the formality 
map $U$.
To reproduce his formula we need to introduce a $(2n+m-2)$--dimensional configuration space
$C^+_{\{p_1,...,p_n\},\{q_1,...,q_m\}}$. If $\cal H$ denotes the upper half-plane then
$C^+_{\{p_1,...,p_n\},\{q_1,...,q_m\}}$ is a quotient of
$$\left\{(p_1,...,p_n;q_1,..,q_m)| p_i\in{\cal H},~q_j\in\BR,~p_{i_1}\neq
p_{i_2}~\hbox{for}~i_1\neq i_2,~q_1<...<q_m\right\}$$
by the action of the group $G^{(1)}=\{z\mapsto az+b|a,b\in \BR, a>0\}$ of orientation-preserving 
affine transformations of the real line.

Then
\beq
U_n = \sum_{m \geq 0}\sum_{\Gamma \in G_{n,m}} w_{\Gamma}B_{\Gamma}.\label{explfor}
\eeq
Here the second summation goes over all oriented admissible graphs with $n$ vertices 
$p_1,...,p_n$ of the first type and $m$ vertices $q_1,..,q_m$ of the second type. 

The rules are:
There are no 
outgoing edges from the second type vertices. There are $k_1, k_2,...,k_n$ edges starting in the 
the first type vertices which are ending in either first type vertices again or in the second 
type vertices. There are no edges starting and ending in the same vertex. The vertices and edges
are enumerated in a fashion compatible with the orientation,
the edges starting
at the first type vertex $p_i$ are labeled by numbers $k_1+k_2+...+k_{j-1}+1,...,
k_1+k_2+k_j$. We denote 
$$Star(p_j)=\{\overrightarrow{p_ja_1},...,\overrightarrow{p_ja_{k_j}}\}\qquad
\overrightarrow{v}_{k_1+...+k_{j-1}+i}=\overrightarrow{p_ja_i}.$$

The weight $w_\Gamma$ of the oriented graph $\Gamma$ is defined
as an integral over the $(2n+~m-2)$-dimensional configuration space $C^+_{\{p_1,...,p_n\},
\{q_1,...,q_m\}}$ 
\beq
w_{\Gamma}=\int\frac{1}{(2\pi)^{\sum k_i}k_1!...k_n!}d\phi_{\overrightarrow{v}_1}
\wedge ...\wedge d\phi_{\overrightarrow{v}_{k_1+...+k_n}},\label{w}
\eeq
where 
\beq
\phi_{\overrightarrow{p_ja}}=Arg\left({a-p_j\over a-\overline{p_j}}\right).
\eeq
If $\alpha_1$ is of degree $k_1-1$, $\alpha_2$ is of degree $k_2-1$,...,
$\alpha_n$ is of degree $k_n-1$, then
\begin{eqnarray}
 \lefteqn{B_\Gamma(\alpha_1,...,\alpha_n)(f_1,f_2,...,f_m)}\nonumber \\[4pt]
&&=\sum D_{p_1}
\alpha_1^{i_1i_2...i_{k_1}}...D_{p_n}\alpha_n^{i_{k_1+...k_{n-
1}+1}...i_{k_1+...+k_n}}D_{q_1}f_1...D_{q_m}f_m,
\end{eqnarray}
where
\beq
D_a=\prod_{l,\overrightarrow{v_l}=\overrightarrow{.a}}\partial_{i_l}
\eeq
and the summation runs over repeated indices $i_j$. 
Finally
\beq
U_n =\sum U_{(k_1,..,k_n)}.
\eeq
The weight $w_{\Gamma}$ is nonzero only if the degree of the polydifferential operator and 
the overall degree of the polyvector fields match as  
\beq
m=2-n+\sum_{i=1}^n (k_i-1).\label{match}
\eeq
Only in this case the degree of the form in (\ref{w}) matches the dimension of the 
configuration space $C^+_{\{p_1,...,p_n\},\{q_1,...,q_m\}}$.
The construction can be globalized to any (formal) Poisson manifold~\cite{Kontsevich}.

To make a relation with the deformation quantization a formal parameter, the 
Planck constant $\hbar$, has to be introduced. If $\alpha$ is a two-tensor, then
by the condition (\ref{match}) $U_n(\alpha,...,\alpha)$ is a bidifferential operator 
for every $n$, i.e., it acts on two functions. 
If moreover $\alpha$ is a Poisson tensor then the Kontsevich 
star product $\star$ is defined 
for $f$ and $g$, two smooth functions
on $M$, as
\beq
f\star g=\sum_{n\geq0} \frac{\hbar^n}{n!}U_n(\alpha,...,\alpha)(f,g) \label{star}. 
\eeq  
The associativity of such a star product follows from  the formality equation. 
If we set in (\ref{formality}) $\alpha_i=\alpha$, for $i=1,..,n$ and take into account the 
Jacobi identity $[\alpha,\alpha]$ and the condition (\ref{match}) 
we see that (\ref{formality})
is equivalent to the $\hbar^n$--order term of the associativity condition 
for $\star$.

There are some other consequences from the formality theorem, which will be useful later
\cite{Manchon}. Here we present them in form convenient for the further use.
We adopt the following notation; for any vector field $\xi$ on 
the Poisson manifold $(M,\alpha)$
we denote as 
$\delta_{\xi}$ the following formal series in $\hbar$
\beq
\delta_{\xi}=\sum_{n\geq1}\frac{\hbar^{n-1}}{(n-1)!}U_n(\xi,\alpha,...,\alpha)\label{vectorlift}.
\eeq
{}From (\ref{match}) we see that this is a differential operator. Its first term equals to $\xi$.
Let us apply the formality equation in the case where $\alpha_1=\xi$ and 
$\alpha_2=...=\alpha_n=\alpha$. The matching condition (\ref{match}) says that in this case
RHS of the formality equation is a bidifferential operator; so let us act both sides of (\ref{formality})
on two functions $f$ and $g$.
Using (\ref{match}) also in evaluating the LHS we find:
\beq
\delta_{\xi}(f\star g)-\delta_{\xi}(f)\star g - f\star\delta_{\xi}(g)=\sum_{n\geq 2}
\frac{\hbar^{n-1}}{(n-2)!}U_{n-1}([\xi,\alpha],\alpha,..\alpha)(f,g). \label{Liepr}
\eeq 
In particular we see that vector fields preserving the Poisson bracket are lifted via $\delta$
to derivations of the Kontsevich's star product.

Further, if we use formality equation and the matching condition in the case,
$\alpha_1=f$, $\alpha_2=...=\alpha_n=\alpha$, we see that the RHS of (\ref{formality})
is a differential operator. Acting with both sides of the formality equation on a function $g$
we get
\beq
\frac{1}{\hbar}[\hat{f},g]_{\star}=\delta_{[\alpha,f]}(g),\label{hat}
\eeq
where the function $\hat{f}$ is given by a formal power series
\beq
\hat{f}=\sum_{n\geq1}\frac{\hbar^{n-1}}{(n-1)!}U_n(f,\alpha,...,\alpha),\label{functionlift}
\eeq
starting with $f$. So the Hamiltonian vector fields are lifted to the inner derivations of the star product.

Let us mention that \cite{CattaneoFelder} gives a very nice field theoretical 
interpretation of the formality map. 

\section{Quantum}
With the help of the formality theorem literally everything in the section 3 
can be 
quantized. To be consistent with physics conventions used in the Section 2 we have 
to replace $\hbar$ by $i\hbar$ everywhere.
Let us quantize the Poisson structure $\theta(t)$ (\ref{sol}) on $M$ for any $t\in[0,1]$ via 
Kontsevich's
deformation quantization (\ref{star})
\beq
f\star_t g=\sum_{n\geq0} \frac{(i\hbar)^n}{n!}U_n(\theta(t),...,\theta(t))(f,g). 
\eeq  
That way we get a star product $\star_t$ for any $t$ .
For any two $t$-independent functions $f$ and $g$ on $M$ we can take the $t$-derivative
of $f\star_t g$. Then equations (\ref{Lie}) and (\ref{Liepr}) give the quantum version
of (\ref{der})
\beq 
\partial_t (f\star_t g) + \delta_{\chi(t)}(f\star_t g)-\delta_{\chi(t)}(f)\star_t g
- f\star_t \delta_{\chi(t)}(g)=0, \label{qder}
\eeq
with $\delta_{\chi(t)}$ given by (\ref{vectorlift})
\beq
\delta_{\chi(t)}=\sum_{n\geq1}\frac{(i\hbar)^{n-1}}{(n-1)!}U_n(\theta(t)(A),\theta(t),...,\theta(t)).
\eeq

This means that we can relate the star products $\star_t$ at $\star_{t'}$ at two different time 
instants by ${\cal D}_{\scriptsize tt'}$, the ``flow" of $\delta_{\chi(t)}$
(or the ``quantum flow" of $\chi(t)$).
Particularly for $t=0$ and $t'=1$ we have
\beq
{\cal D} = \left.e^{\partial_t + \delta_{\chi(t)}}e^{-\partial_t}\right |_{t=0}=\left.e^{-\partial_t} 
e^{\delta_{\chi(t+1)} +\partial_t} \right |_{t=0}.
\eeq
Let us note that $\cal D$ is a composition $D\circ\rho^*$ of the classical flow $\rho^*$ and 
a gauge equivalence $D={\cal D}\circ (\rho^*)^{-1}$ of the star product $\star_{\rho}$ obtained from $\star'$ by simple 
action of $\rho^*$ and the star product $\star$.

Finally the gauge transformation (\ref{gauge}) is quantized with the help of (\ref{tilde}),
(\ref{hat}) and (\ref{functionlift})
as
\beq
f \stackrel{(\lambda)}{\mapsto} f + \frac{1}{i\hbar}[\hat{\tilde \lambda},f]_{\star}, \label{hattilde}
\eeq
where 
\beq
\hat{\tilde\lambda}=\sum_{n\geq1}\frac{(i\hbar)^{n-1}}{(n-1)!}U_n(\tilde \lambda,\theta,...,\theta)
\eeq 
and $\tilde{\lambda}$ is obtained as explained in section 3 from the condition
\beq
e^{[\theta, \tilde \lambda]}=\left.e^{\partial_t + \chi_{\lambda}(t)}e^{-\partial_t - 
\chi(t)}\right|_{t=0}
\eeq
using the B-C-H formula.

The rest is trivial. We write similarly to (\ref{xplusa})
\beq
{\cal D} (x^i)=\left.e^{\partial_t + \delta_{\chi(t)}}x^i \right |_{t=0} = x^i + \hat{A}^i.
\eeq
Again $\hat A$ depends as a formal power series in $\theta$ on $A$.
Explicitly we have
\beq
\hat{A}^i = (\left.\mbox{exp}(\partial_t + \delta_{\theta^{ij}(t)A_i\partial_j})-1)x^i\right |_{t=0} \label{QSW}
\eeq

If we act by the infinitesimal gauge transformation (\ref{gauge}) on $A$,
this induces now the action of the inner derivation (\ref{hattilde}) on ${\cal D}(x^i)$,
which in turn induces a map on $\hat A$
given by
\beq
\hat{A}^i \stackrel{(\lambda)}{\mapsto} \hat{A}^i + \frac{1}{i\hbar}[\hat {\tilde \lambda},x^i]_{\star}
+ \frac{1}{i\hbar}[\hat{\tilde \lambda}, \hat{A}^i]_{\star}.
\eeq

So (\ref{QSW}) gives indeed the desired SW map to all orders in $\theta$
in the case of a general Poisson manifold.
Moreover, using Kontsevich's construction of the formality map $U$ 
as described in section 4, we can find explicit expressions in local coordinates 
for $\hat A$ and $\hat{\tilde \lambda}$ to any order in $\theta$.

To conform completely to the conventions we adopted in section 2, we could have 
taken $-A$ and $-\lambda$
as the actual classical gauge field and gauge parameter.

\section*{Acknowledgments}

We would like to thank S.\ Theisen for encouraging discussions. B.J.\ thanks
the Alexander-von-Humboldt-Stiftung for support and J. Donin, P. Bressler and
S. Merkulov for discussions.

\end{document}